\documentclass[tightenlines,
twocolumn,a4paper,
showpacs,nofootinbib,aps]{revtex4}

\usepackage{epsfig}
\usepackage{axodraw}

\usepackage{amssymb}

\newcommand{\bea}{\begin{eqnarray}}
\newcommand{\eea}{\end{eqnarray}}
\newcommand{\bean}{\begin{eqnarray*}}
\newcommand{\eean}{\end{eqnarray*}}
\newcommand{\nn}{\nonumber \\}

\def\W #1{\widetilde{#1}}

\def\braket#1{\left\langle #1 \right\rangle}

\def\gb #1{ \left\langle #1 \right]}

\def\eps{\epsilon}

\def\Label#1{\label{#1}%
  \smash{\hbox to0pt{\raise1ex\hbox{\tiny[#1]}\hss}}}

\begin{document}
\begin{titlepage}

\begin{flushright}
\vbox{
\begin{tabular}{l}
ITFA-2006-35\\
ZU-TH 21/06
\end{tabular}
}
\end{flushright}

\title{ D-dimensional unitarity cut method}

\author{
Charalampos Anastasiou$^{a}$,  Ruth Britto$^{b}$,
Bo Feng$^{c,d}$,\\  
Zoltan Kunszt$^{a}$ and Pierpaolo Mastrolia$^{e}$}
\affiliation{
$^a$Institute of Theoretical Physics, ETH Zurich, 8093 Zurich, Switzerland \\
$^b$Institute for Theoretical Physics, University of Amsterdam
Valckenierstraat 65, 1018 XE Amsterdam, The Netherlands\\
$^c$Blackett Laboratory \& The Institute for Mathematical Sciences, Imperial College, London, SW7 2AZ, UK\\
$^d$ Center of Mathematical Science, Zhejiang University, Hangzhou, China\\
$^e$Institute for Theoretical Physics, University of Zurich,
8057 Zurich, Switzerland
}

\begin{abstract}
We develop a unitarity method to compute one-loop amplitudes with
massless propagators in $d=4-2\epsilon$ dimensions. We compute
double cuts of the loop amplitudes via a decomposition into a four-dimensional and a $-2\epsilon$-dimensional integration. The four-dimensional
integration is performed using spinor integration or other efficient
techniques. The remaining integral in $-2\epsilon$ dimensions is
cast in terms of  bubble, triangle, box, and pentagon master
integrals using  dimensional shift identities. The method yields
results valid for arbitrary values of $\epsilon$.
\end{abstract}

\maketitle

\thispagestyle{empty}
\end{titlepage}

\section{Introduction}
In modern collider experiments complex  events with multi-jets,
vector bosons and jets, top quarks  and  jets, etc.\  are
frequently produced. Their  quantitative theoretical
description  requires   cross-sections   calculated  at the one-loop level
and  even beyond.
There exist mature techniques solving
all conceptual problems  which arise in one-loop computations. However,
calculating  one-loop multi-leg amplitudes with standard methods
is  tedious, due to the large number of Feynman diagrams and
the algebraic complexity  of tensor  reduction.
In recent years, new attempts are being made to replace or
improve traditional approaches with more efficient and better automated
methods. Significant progress can be expected with the advent of new
powerful techniques.

Unitarity cuts  of loop amplitudes have been introduced as an
efficient tool to calculate QCD amplitudes~\cite{Bern:1994zx}. A new
four-dimensional unitarity method was developed recently~\cite{Britto:2004nc,Britto:2005ha},  building on
techniques
inspired by twistor space geometry~\cite{Witten:2003nn,Cachazo:2004kj,Cachazo:2004by,Brandhuber:2004yw,Cachazo:2004zb,Bena:2004xu,Cachazo:2004dr,Britto:2004nj}.
The phase-space integration  is carried out explicitly
in terms of spinors. The result  is easily mapped to bubble, triangle, and box master
integrals using analyticity properties. Many, mostly supersymmetric,
amplitudes can be 
reconstructed fully 
with this
technique~\cite{Britto:2004nc,Britto:2005ha,Britto:2006sj} and other
unitarity methods in four  dimensions. However, the mapping to
master integrals is in general incomplete, since rational
contributions arising from multiplying $1/\eps$ poles  of master
integrals with ${{\cal O}}(\epsilon)$ coefficients are not accounted
for.

New methods to compute the rational parts separately were introduced
recently. They compute these terms by either
developing~\cite{Bern:2005hs,Bern:2005ji,Bern:2005cq,Berger:2006ci,Berger:2006vq}
recursion relations for amplitudes~\cite{Britto:2004ap,Britto:2005fq}, or by using
specialized diagrammatic
reductions~\cite{Xiao:2006vr,Su:2006vs,Xiao:2006vt,Binoth:2006hk}. As a result,
for example, short analytic formulas are now available for  all the
one-loop six gluon QCD helicity amplitudes.

It was recognized long ago~\cite{vanNeerven:1985xr}
that one can reconstruct the full amplitudes  from
unitarity cuts in $d=4-2\epsilon$ dimensions. 
A complete method for one-loop calculations 
was developed in the pioneering work of Bern 
{\em et al.}~\cite{Bern:1995db,Bern:1996je,Bern:1996ja}, and it was recently  
re-examined in~\cite{Brandhuber:2005jw}.
However, the calculation  of general unitarity cuts
remains formidable. While it is simpler than a direct Feynman
graph evaluation, eventually, one resorts to traditional  reduction
methods to complete their computation.

In this Letter, we develop an efficient $d$-dimensional unitary cut
method, reducing  one-loop  amplitudes to master integrals
for arbitrary values of the dimension parameter.
We can read out the coefficient of the master integrals without fully
carrying out the $d$-dimensional phase space integrals.
Only a four dimensional integration is explicitly required;
we show how to perform this using spinor integration for
{\it light-like} momenta~\cite{Cachazo:2004dr,Britto:2004nj,Britto:2004nc,Britto:2005ha}.
A remaining integral, which gives rise to the $\epsilon$-dependence
of the cut-amplitude, is mapped to phase-space integrals in
$4+2n-2\epsilon$ dimensions, where $n$ is a positive integer. With
recursive dimensional shift identities, similar to the ones in loop
integration, we reduce the cut-amplitude in terms of bubble, triangle, box
and pentagon cut master integrals in $4-2\epsilon$ dimensions.
The reduction is valid for an arbitrary number of dimensions.
Expanding  in $\epsilon$, we can obtain both the (poly)logarithmic and
rational part of the amplitude at ${\cal O}(\epsilon^0)$ and higher; 
these contributions are required in cross-sections beyond the 
next-to-leading order in the relevant coupling strength.

The core part of our method is the four-dimensional integration,
where we have primarily used spinor integration. In
\cite{Britto:2005ha,Britto:2006sj} it was demonstrated that this
method is very efficient and  yields compact results for the
cut-constructible part of multi-leg QCD amplitudes. In our
$d$-dimensional unitarity case the four-dimensional integrand depends
on an additional mass parameter. While the size of expressions is
larger, spinor integration works efficiently by preserving
gauge invariance at intermediate stages of the computation. The
final results remain compact.

\section{Reduction to master integrals}

We consider one-loop amplitudes
with massless internal propagators
in the four-dimensional helicity  (FDH)
scheme; all external momenta are in four dimensions and the loop
momentum in $d=4-2\epsilon$.  We shall reduce double cuts of the amplitude
to master integrals, for arbitrary  values of $\epsilon$.

The basic quantity that  we require  is a generic double cut  of the
amplitude in $4-2\epsilon$ dimensions:
\begin{equation}
{\cal M} = \int d^{4-2\epsilon} p
\delta\left(p^2\right) \delta\left((K-p)^2\right)
{\cal A}_L(p) {\cal A}_R(p),
\label{eq:doublecut}
\end{equation}
where ${\cal A}_{L,R}$ are tree amplitudes, and $K$ is the sum of
the momenta of the cut  propagators. Since  external momenta are
in four dimensions, we can decompose the loop momentum as 
$p=\W \ell+\vec{\mu}, $ where $\W
\ell$ is $4$-dimensional and $\vec{\mu}$ is $(-2\eps)$-dimensional~\cite{Bern:1995db,Bern:1996je,Bern:1996ja}. One can immediately perform the angular integrations for
$\vec{\mu}$, yielding:
\begin{eqnarray}
&&
{\cal M} = \frac{\pi^{-\epsilon}}{\Gamma(-\epsilon)}
\int d\mu^2 {(\mu^2)}^{-1-\epsilon}  \nonumber \\
&&   \times
\int d^{4} \W {\ell}
\delta
\left(
{\W{\ell}}^2-\mu^2
\right)
\delta\left((K-\W{\ell})^2 -\mu^2\right)
\nonumber \\
&&
\qquad
{\cal A}_L\left(\W{\ell}+\vec{\mu}\right)
{\cal A}_R\left(\W{\ell}+\vec{\mu}\right).
\label{eq:doublecut4dim}
\end{eqnarray}
The unitarity cut integral of massless particles living in $d$ dimensions
is decomposed  into a unitary cut integral of massive particles
in four dimensions, and an integral over the mass parameter regularized
with $\epsilon$.

We now perform the integration over $\W{\ell}$. 
Given the virtues of
spinor
integration,
it  is desirable to employ it here. However, the method  is
formulated for phase-space integrations of light-like particles and,
at first sight, is  not applicable to our case.  We can find a
generalization to the phase-space of massive particles,  if  we
decompose the momentum $\tilde{\ell}$ into a linear combination of a
light-like vector and the time-like cut-momentum $K$: $\W \ell=
\ell+z K$, with $\ell^2=0$.
The massive phase space integral turns into massless:
\begin{eqnarray}
&& \int d^4 \W \ell \delta(\W\ell^2-\mu^2) \delta((\W \ell-K)^2-\mu^2)
\nonumber  \\
&& \qquad \to \int dz  (1-2z) K^2\delta(z(1-z)K^2-\mu^2) \nonumber
\\ && \qquad\int d^4\ell \delta^+(\ell^2) \delta( (1-2z)K^2-2K\cdot \ell).
\end{eqnarray}
%
The last line is the familiar phase space integration for
two massless cut propagators;
 the  only difference  is
the factor $(1-2z)$ appearing in the second
delta function. The $z$-integral is trivially performed
using the delta-function that is independent of $\ell$. Thus we
get $z={(1-\sqrt{1-u})/2}$, where $u={4\mu^2/K^2}\in [0,1]$.

Following~\cite{Cachazo:2004kj,Cachazo:2004dr,Britto:2005ha}
we transform into spinor variables, so that $ \ell^{a\dot{a}} = t
\lambda^{a} \tilde{\lambda}^{\dot{a}}$. The phase-space
measure, up to an overall normalization factor, becomes:
\begin{eqnarray}
&& \int du u^{-1-\epsilon}
\int 
\braket{\lambda \, d\lambda} [\lambda
\, d \lambda] \nonumber \\
&& 
\int_0^\infty 
t dt \delta \left( \sqrt{1-u} K^2+t
\gb{\lambda|K|\lambda} \right) \label{eq:psmeasure}
\end{eqnarray}
The spinor integration can be carried out easily. The basic steps
involve the application of Schouten identities in order to
eliminate $\tilde{\lambda}^{\dot{a}}$ from the numerator of the
integrand, and to locate  holomorphic anomalies, reading out the
result of the integration as  a finite sum of residues. We refer the
reader to~\cite{Britto:2005ha,Britto:2006sj}
for a detailed description of the technique.

After spinor integration, we are left with a single integral over $u$,
\begin{eqnarray}
{\cal M} = \int_0^1 du u^{-1-\eps}\sum_{i} f_i(u) {\cal L}_i(u),
~\label{C-gen}
\end{eqnarray}
where the coefficients $f_i(u)$ are rational functions of $u$.
The  functions ${\cal L}_i$ are combinations
of logarithmic and square root functions with characteristic analyticity
properties; they correspond to the analytic expressions of
massive cut master integrals (bubbles, triangles, and boxes) in four
dimensions.

We can express the cut amplitude ${\cal M}$ in terms of master integrals in
$4-2\epsilon$ dimensions without explicitly performing the
integration in Eq.~\ref{C-gen}. Many coefficients $f_i(u)$ are
simple polynomials in $u$.  All such terms are easily identified
as one-loop master integrals in dimensions shifted by an even
number $2n$.  Schematically, bubble, triangle, and box
master integrals emerge in the form:
\begin{eqnarray}
&& {\rm Bub}^{(n)} =  \int_0^1 du u^{-1-\eps} u^n
\sqrt{1-u} \label{eq:bubmaster}
\\ \nn
&& {\rm Tri}^{(n)} = \int_0^1 du u^{-1-\eps} u^n\ln
\left( {Z +\sqrt{1-u}\over Z-\sqrt{1-u}
}\right) \nonumber \\
\label{eq:trimaster}
\\ \nn
&& {\rm Box}^{(n)} = \int_0^1 du u^{-1-\eps} {u^n\over
\sqrt{B - A u}} \times
\nonumber \\
&& \ln \left( {D - C u- \sqrt{1-u}\sqrt{
B - A u}\over D - C u+ \sqrt{1-u}\sqrt{
B - A u}}\right)
\label{eq:boxmaster}
\end{eqnarray}
where $Z^2, A, B, C, D$ are rational functions of kinematical
invariants of  the external momenta. Details of the exact
expressions will be given in a forthcoming publication.
While mapping to $\epsilon$-dependent master integrals,  a term of the
form $u^n$ is always absorbed into the measure factor $u^{-1-\eps}$, producing
a dimensional shift.

After partial fractioning and identifying all 
$(4-2\epsilon+2n)-$dimensional
bubble, triangle, and box master integrals, a few coefficients $f_i$
are yet not mapped to any master integral. These coefficients have a
$u$-dependent monomial in the denominator, and multiply logarithms
originating from box master integrals in four dimensions. They are
related to the pentagon scalar integral, which can be expressed as
a sum of box integrals in four dimensions plus some term in higher
order of $\eps$.  However, it is a master in arbitrary dimensions.
Upon integrating over $u$, the remaining terms  combine to give rise
to pentagon master integrals in $4-2\epsilon$ dimensions.

As a last step, we reduce the master integrals in $4-2\epsilon + 2n$
dimensions, to master integrals in $4-2\epsilon$ dimensions. We can
derive compact dimensional shift identities for the phase-space
master integrals from the representations in
Eqs.~\ref{eq:bubmaster}-\ref{eq:boxmaster} using integration by
parts. These identities are equivalent to dimensional shift
identities for loop 
integrals~\cite{Bern:1992em,Bern:1993kr,Tarasov:1996br}.  We
will give their explicit analytic form and a simple derivation in a
forthcoming publication. Here we just present the results. 
\begin{eqnarray}
&&\hspace{-0.5cm}  {\rm Bub}^{(n)} = F^{(n)}_{2 \to 2} {\rm Bub}^{(0)}
\nonumber \\
&& \hspace{-0.5cm} {\rm Tri}^{(n)}(Z) = F^{(n)}_{3 \to 3}(Z) {\rm Tri}^{(0)}(Z)
+ F^{(n)}_{3\to 2}(Z) {\rm Bub}^{(0)}
\nonumber \\
&&\hspace{-0.5cm} {\rm Box}^{(n)}
 = F^{(n)}_{4 \to 4} {\rm Box}^{(0)}
+ \bigg \{ 
F^{(n)}_{4 \to 3}(Z_1) {\rm Tri}^{(0)}(Z_1) 
\nonumber \\
&& \hspace{0.7cm}
+ F^{(n)}_{4\to 2}(Z_1) {\rm Bub}^{(0)}
+ (Z_1 \leftrightarrow Z_2) \bigg\}
\nonumber \\
&&\hspace{-0.5cm}  F^{(n)}_{2 \to 2} =  {(-\eps)_{3 \over 2} \over (n-\eps)_{3 \over 2}}, \quad F^{(n)}_{3 \to 3} =  {-\eps\over n-\eps} (1-Z^2)^n,
\nonumber \\
&& \hspace{-0.5cm}
 F^{(n)}_{4 \to 4} = { (-\eps)_{1 \over 2} \over (n-\eps)_{1 \over 2}}  
\left(  {B \over A} \right)^n,
\nonumber \\
&&\hspace{-0.5cm}  F^{(n)}_{3 \to 2} =  {(-\eps)_{3 \over 2} \over n-\eps}  \sum_{k=1}^n {{2 Z}(1-Z^2)^{n-k} \over (k-\eps)_{1 \over 2}}
\nonumber \\
&&\hspace{-0.5cm} F^{(n)}_{4 \to j} ={D + (Z^2-1) C \over (n-\eps)_{1 \over 2} Z A } \sum_{k=1}^n
\left(  {B \over A} \right)^{n-k} \hspace{-0.4cm}{F^{(k-1)}_{3 \to j}\over (k-1/2-\eps)_{1 \over 2}}
\label{eq:reduction}
\end{eqnarray}
Here $(x)_n = \Gamma(x+n)/\Gamma(x)$, and $j=2,3$. $Z_1,Z_2$ correspond to the 
two possible cut-triangles  obtained by pinching the uncut propagators of the box.

In this Letter, we have limited our work to one-loop amplitudes
with massless internal propagators. However, the method can
be extended to the massive case. The  spinor integration method
is already adapted to massive phase-space. The only remaining issue is
to find the reduction coefficient of master integrals with only one
loop propagator; these vanish when a two-particle phase-space  is
considered. However, such terms are significantly constrained and often
fully determined from the known ultraviolet and infrared behavior of
one-loop amplitudes~\cite{Bern:1995db}. We will investigate this
issue further in a future publication.

\section{Alternatives to spinor integration}

We have seen that our calculations can be
divided into two steps: the four-dimensional-massive cut-integration and
the dimensional shift. 
We have the flexibility of using alternative methods for the four-dimensional integration to compute the coefficients $f_i(u)$ in
Eq.~\ref{C-gen}. Following a more traditional approach, we could
apply the phase-space reduction methods of~\cite{Anastasiou:2002yz}.
The integrals we consider here are free from both infrared and
ultraviolet singularities, and no dimensional regulator is required.
In precisely four dimensions, the reduction is much less tedious
than in arbitrary dimensions. In many complicated cases it can be
performed analytically, and in all cases of practical interest it
can also be executed
numerically~\cite{Laporta:2001dd,Anastasiou:2004vj,Giele:2004iy,Giele:2004ub,Anastasiou:2005cb}.
This technique provides a valuable cross-check on our results
with spinor integration.

Another appealing idea has appeared recently in the literature.
Ossola, Papadopoulos and Pittau  (OPP) introduced a purely algebraic
procedure to compute master integral coefficients at the integrand
level~\cite{Ossola:2006us}.
One can adapt
the same technique for the four dimensional phase-space integration
over cut amplitudes. As an ingredient of the $d$-dimensional unitarity
method, it should be a very efficient tool to analytically compute 
one-loop amplitudes in arbitrary dimensions.

OPP investigated the most general analytic form of one-loop amplitudes
in four dimensions.
The integrands of one-loop amplitudes are decomposed as:
\begin{eqnarray}
A(\W \ell) &= & \sum_{i} ( c_i + \sum_j S_{ij}(\W \ell){b_{ij}} ) I_i(\W
\ell)
~~~\label{OPP-1}
\end{eqnarray}
where $I_i$ are products of propagators corresponding to scalar master integrals, and $c_i, b_{ij}$ are constant coefficients.
The universal terms $S_{ij}(\W{\ell})$ are ``spurious'' and
yield a zero contribution to the amplitude after integration
$\int d^4\W\ell  S_{ij}(\W\ell) I_i(\W \ell)=0$; they are 
known explicitly for all master integrals~\cite{Ossola:2006us}.

With the analytic form of Eq.~\ref{OPP-1} at hand, the coefficients
$c_i$,$b_{ij}$ can be evaluated by computing the integrand
algebraically at sufficiently many values of the loop
momentum and forming a linear system of equations.  The method is
optimized  by choosing values of the loop momenta that correspond
to cuts of the loop amplitude, setting  denominators in the master
integrals to zero. In this way, the system is divided into closed
subsystems for the integrals which survive any specific cut and can
be solved easily.

In sum, we can apply any efficient method, for example the method of
~\cite{Ossola:2006us}, to find the coefficients $f_i(u)$ in
Eq.~\ref{C-gen}~\cite{footnote:1}. Note that, in this way, we can also compute the
contributions to one-propagator master integrals which appear in
amplitudes with massive propagators.

\section{Summary}

In this Letter, we have presented a new unitarity method for the reduction
of one-loop amplitudes to master integrals in arbitrary dimensions.
We have generalized the
method of spinor
integration via the holomorphic anomaly
to massive phase-space integrals.
The method consists of an explicit four-dimensional integration over the
phase-space of double-cut amplitudes, and a remaining integration over
a mass parameter.

 As a cross-check of the four-dimensional integration, one may employ
traditional, numerical and analytic phase-space
reductions.
Recently, an elegant proposal
to compute the reduction coefficients of one-loop amplitudes has
appeared in the literature~\cite{Ossola:2006us}.  
This proposal may also be adopted within our method
in order to perform the four-dimensional integration.

The final integration over a mass parameter is mapped directly to phase-space
master integrals with shifted dimensions. A full reduction to master
integrals in $4-2\epsilon$ is achieved with compact dimensional
shift identities.

We anticipate our method to be useful for a wide spectrum of processes
at colliders.


\section*{Acknowledgments}
CA is supported by the Swiss National Fund under contract 
NF-Projekt 20-105493.
RB is supported by Stichting FOM.  
BF and PM are supported by the Marie-Curie Research 
Training Network under
contracts MRTN-CT-2004-005104 and MEIF-CT-2006-024178.  
RB and BF are grateful for the hospitality of ETH Zurich.
We thank the participants of the HP$^2$ workshop in Zurich, 
September 2006.


\end{document}